\documentstyle[12pt]{article}
\def\beq{\begin{equation}}
\def\eeq{\end{equation}}
\def\bea{\begin{eqnarray}}
\def\eea{\end{eqnarray}}
\def\bq{\begin{quote}}
\def\eq{\end{quote}}

\def\nnb{\nonumber}
\def\ga{\left(}
\def\dr{\right)}
\def\aga{\left\{}
\def\adr{\right\}}

\def\rar{\rightarrow}
\def\nnb{\nonumber}
\def\la{\langle}
\def\ra{\rangle}
\def\nin{\noindent}
\begin{document}
\pagestyle{empty}
\begin{flushright}
PM 95/43\\
\end{flushright}
\begin{center}
\section*{HEAVY FLAVOURS FROM \\QCD SPECTRAL SUM RULES}
\vspace*{1.0cm}
{\bf S. Narison} \\
\vspace{0.3cm}
Laboratoire de Physique Math\'ematique\\
Universit\'e de Montpellier II\\
Place Eug\`ene Bataillon\\
34095 - Montpellier Cedex 05, France\\
\vspace*{2.0cm}
{\bf Abstract} \\ \end{center}
\noindent
We present a summary update of the QCD spectral sum rule (QSSR)
results for the
running and {\it perturbative pole} quark masses, the $f_D$ and
$f_B$
leptonic decay constants, the heavy-to-light and heavy-to-heavy
exclusive transition-form factors. Analytic expressions of
these latter quantities are presented, which give a deeper
understanding
of their $q^2$- and infinite mass-behaviours.
A short comparison of the QSSR results with alternative approaches
is done.
\vspace*{2cm}
\nin
\begin{center}{\it {Talk given at the session on Non-Perturbative
Methods,\\ EPS Conference, Brussels (July 1995)}}
\end{center}
\vspace*{2cm}
\begin{flushleft}
PM 95/43\\
October 1995
\end{flushleft}
\vfill\eject
\setcounter{page}{1}
 \pagestyle{plain}
\section{QCD Spectral Sum Rules (QSSR)}
\nin
QCD spectral sum rule (QSSR) \`a la SVZ \cite{SVZ} (for a recent
review, see e.g. \cite{SNB}) has shown since
15 years, its impressive ability
for describing the complex phenomena of hadronic
physics with the few universal ``fundamental" parameters of the QCD
Lagrangian
(QCD coupling $\alpha_s$, quark masses
and  vacuum condensates built from the quarks
and/or gluon fields), without waiting for a
complete understanding of the confinement problem.
In the example of the two-point correlator:
\beq
\Pi_b(q^2) \equiv i \int d^4x ~e^{iqx} \
\la 0\vert {\cal T}
J_b(x)
\ga J_b(o)\dr ^\dagger \vert 0 \ra ,
\eeq
 associated to the generic hadronic current:
$J_b(x) \equiv \bar q \Gamma b (x)$ of the $q$ and $b$-quarks
($\Gamma$ is
a Dirac matrix which specifies the hadron quantum numbers),
the SVZ-expansion reads:
\bea
\Pi_b (q^2)
&\simeq& \sum_{D=0,2,...}\sum_{dim O=D}
\frac{ C^{(J)}(q^2,M^2_b,\mu)\la O(\mu)\ra}
{\ga M^2_b-q^2 \dr^{D/2}},
\eea
where $\mu$ is an arbitrary scale that separates the long- and
short-distance dynamics; $C^{(J)}$ are the Wilson coefficients
calculable
in perturbative QCD by means of Feynman diagrams techniques;
$\la O \ra$ are
the non-perturbative condensates of dimension $D$ built
 from the quarks or/and gluon
fields ($D=0$
corresponds to the case of the na\"\i ve perturbative
contribution). Owing to gauge invariance, the lowest
dimension condensates that can be formed are the $D=4$
light quark $m_q \la\bar \psi \psi \ra$ and gluon $\la\alpha_s
G^2 \ra$
ones, where the former is fixed by the pion PCAC relation, whilst
the latter
is known to be $(0.07\pm 0.01)$ GeV$^4$ from more recent analysis
of the light \cite{SNL} and heavy quark systems \cite{SNB}.
The
validity of the SVZ-$postulate$ has been understood formally, using
 renormalon
techniques (absorption of the IR renormalon ambiguity into the
definitions
of the condensates, UV renormalon cannot induce
some extra $1/M^2$-terms not included
in the OPE)
\cite{MUELLER,BENEKE} and/or by building  renormalization-invariant
combinations of the condensates (Appendix of \cite{PICH} and
references
therein). The SVZ expansion is phenomenologically confirmed from
the unexpected
accurate determination of the QCD coupling $\alpha_s$ and from
a measurement of the condensates from semi-inclusive
tau decays \cite{PICH,ALFA}.
\nin
 The previous QCD information is transmitted to the data through
the spectral function Im$\Pi_b(t)$
via the K\"allen--Lehmann dispersion relation ($global~duality)$
 obeyed by the hadronic correlators,
which can be improved from the uses of
either a finite number of derivatives and finite values of $q^2$
(moment sum rules):
\beq
{\cal M}^{(n)} \equiv \frac{1}{n!}\frac{\partial^n \Pi_b(q^2)}
{\ga \partial q^2\dr^n} \Bigg{\vert} _{q^2=0}
= \int_{M_b^2}^{\infty} \frac{dt}{t^{n+1}}
{}~\frac{1}{\pi}~ \mbox{Im}  \Pi_b (t),
\eeq
or an infinite number of derivatives and infinite values of $q^2$,
but keeping their ratio fixed as $\tau \equiv n/q^2$
(Laplace or exponential sum rules):
\beq
{\cal L}(\tau,M^2_b)
= \int_{M_b^2}^{\infty} {dt}~\mbox{exp}(-t\tau)
{}~\frac{1}{\pi}~\mbox{Im} \Pi_b(t),
\eeq
for $m_q=0$.
Non-relativistic versions of these two sum rules
are convenient quantities to work with, in the large-quark-mass
limit, after introducing the non-relativistic
variables $E$ and $\tau_N$:
$t \equiv (E+M_b)^2$ and $\tau_N \sim M_b\tau .$
In the previous sum rules,
the weight factors enhance the
contribution of the lowest ground-state meson to the spectral
integral,
such that, the simple duality ansatz parametrization:
$
``{one~narrow~resonance}"+
 ``{QCD~ continuum}"$, from a threshold $t_c$,
gives a very good description of the spectral integral.
The previous na\"{\i}ve parametrization
has been tested successfully in the light-quark channel from the
$e^+e^- \rar$
$I=1$ hadron data and in the heavy-quark ones from the
$e^+e^- \rar \psi$ or $\Upsilon$ data, within a good
accuracy.
%
In principle, the pairs $(n,t_c)$, $(\tau,t_c)$ are free external
parameters in the analysis, so that the optimal result should be
insensitive to their variations. {\it Stability criteria}, which
are equivalent
to the variational method, state that the best results should
be obtained at the minimas or at the inflexion points in $n$ or
$\tau$,
while stability in $t_c$ is useful to control the sensitivity of
the
result in the changes of $t_c$-values. To these stability criteria
can be
added constraints from local duality FESR, which
correlate the $t_c$-values to those of the ground state mass and
coupling \cite{FESR}. Stability criteria have also been tested in
models such as the
harmonic oscillator  \cite{BELL}, where the exact and approximate
solutions are known.
\nin
 However, though
{\it I would personally expect} that the $true$ result is obtained
near the beginning
of the $t_c$-stability region \footnote{
This value of $t_c$
is about the one fixed by FESR duality constraints.
In this region, the
result is certainly insensitive to the form of the continuum
model.} despite the fact that in some cases
this value of
$t_c$ is larger than the phenomenological guessed position of the
next radial
excitation, one can $fairly$ state that the {\it most conservative
optimization criteria}, which include various types of
optimizations
in the literature, are the one obtained in the region, starting
from
the beginning of $\tau / n$ stability \footnote{This corresponds
in most
of the cases to the so-called plateau often discussed in the
literature,
but in my opinion, the interpretation of this plateau as a
sign of a good continuum model is not sufficient, in the sense
that the flatness of the
 curve extends in the uninteresting high-energy region where the
properties of the ground state are lost.},
until the beginning of the $t_c$
stability.
One can  $a ~posteriori$ check that, at the stability
point, where we have an equilibrium between the continuum and
non-perturbative
contributions, which are both small,
the OPE is still convergent and the expansion certainly makes
sense.
The results which will be quoted below have been obtained within
the previous $stability~criteria$.
\section{The heavy-quark-mass values}
\nin
Many efforts have been devoted to the study of the quark masses
\cite{PDG}.
Using the present $world~average$
value $\alpha_s(M_Z)=0.118 \pm 0.006$ \cite{BETHKE}, the
$first~direct$
determination of the running mass to two loops, from the $\Psi$
and
$\Upsilon$ systems, is \cite{SNM}:
\bea
\overline{m}_c(M^{PT2}_c) &=& (1.23 ^{+ 0.02}_{-0.04}\pm 0.03
)~ \mbox{GeV} \nnb \\
\overline{m}_b(M^{PT2}_b)
&=& (4.23 ^{+ 0.03}_{-0.04}\pm 0.02)~ \mbox{GeV},
\eea
where the errors are respectively due to $\alpha_s$ and to the
gluon condensate. Using the relation \cite{SNM2}:
\beq
M_Q
=\overline{m}_Q(M_Q^2)\aga 1+\frac{4}{3}\ga\frac{\alpha_s}{\pi}\dr
 +K_Q\ga\frac{\alpha_s}{\pi}\dr^2 \adr ,
\eeq
where $K_b \simeq 12.4,~K_c \simeq 13.3$ \cite{SCHILCHER},
one can transform this result into the {\it perturbative
pole} mass and obtain, to two-:
\bea
M^{PT2}_c & = & (1.42 \pm 0.03
)~ \mbox{GeV} \nnb \\
M^{PT2}_b
&= &  (4.62 \pm 0.02)~ \mbox{GeV}~,
\eea
and three-loop accuracy:
\bea
M^{PT3}_c  & =&  (1.62 \pm 0.07 \pm 0.03
)~ \mbox{GeV} \nnb \\
M^{PT3}_b
&=&  (4.87 \pm 0.05 \pm 0.02)~ \mbox{GeV}.
\eea
It is informative to compare these values with the ones of the
pole
masses from HQET \cite{NEU,SNH} and non-relativistic sum rules to
two loops \cite{SNM}:
\bea
M^{NR}_c & =& (1.45^{+ 0.04}_{-0.03} \pm 0.03
)~ \mbox{GeV} \nnb \\
M^{NR}_b
& =&  (4.69 _{-0.01}^{+0.02}\pm 0.02)~ \mbox{GeV},
\eea
where one might interpret the small mass difference
of about 70 MeV as the
size of the renormalon effect into the pole mass. Indeed, an
explicit
resummation of the leading
$(\beta\alpha_s)^n$ terms increases the previous two-loop estimate
by about 100$\sim$200 MeV \cite{BENEKE}. It also indicates
that the three-loop value in Eq. (8) already
gives a good estimate of the pole mass to all orders of PT. Eq. (8)
compares quite well with the $dressed$ mass
$
  M_b^{nr}= (4.94\pm 0.10 \pm 0.03)~\mbox{ GeV},
$
obtained from
a non-relativistic higher-order Balmer formula based on a $\bar bb
$ Coulomb potential \cite{YND} and with lattice calculations
\cite{LATTICEM}. One can also
use the previous results, in order to deduce the
(non)-relativistic pole mass-difference
of the $b$- and $c$-quarks both evaluated at $M_b$:
\beq
M_b-M_c\Big{|}_{p^2=M^2_b} = (3.54 \pm 0.05)~ \mbox{GeV}.
\eeq
Finally, the $lesson$ which we can learn from the previous
discussion is
that one should be very careful in using
 the $numerical~value$ of the quark mass. Indeed, for
consistency, one should first
understand the definition of the mass used in the analysis and
know
to what loop-accuracy the analysis is done (for a recent
compilation of the running quark masses, see e.g. \cite
{RODRIGO}.).
\section{The decay constants and the $B_B$-parameter}
\nin
The decay constants $f_P$ of a pseudoscalar meson $P$ are defined
as:
\beq
(m_q+M_Q)\la 0\vert \bar q (i\gamma_5)Q \vert P\ra
 \equiv \sqrt{2} M^2_P f_P,
\eeq
where in this normalization $f_\pi = 93.3$ MeV. A lot of efforts
have also
been done for the estimate of these decay constants \cite{SNB}.
However, the
most pertinent result has been obtained, {\it for the first time},
in \cite{SN3}:
\beq
f_D\approx f_B \simeq 1.4f_\pi~,
\eeq
which clearly shows a violation of the Infinite Mass Effective
Theory (IMET)
$1/\sqrt{M_b}$-scaling law. Several obscure and unjustified
criticisms have been adressed
later on in order to discredit such a result, but a numerical
estimate of
the $1/M_b$-corrections complemented by the lattice results
and by
the analytic estimate from HQET and semi-local duality has
provided a better
understanding of the $unexpected$ result in Eq. (12). Indeed,
using the present update $best$
estimate from the Laplace sum rule \cite{SN2}:
\bea
f_D &\simeq& (1.35 \pm 0.04\pm 0.06)f_\pi\nnb\\
f_B &\simeq& (1.49\pm 0.06\pm 0.05)f_\pi~,
\eea
 consistent with the value of the relativistic two-loop pole mass
given previously, and using
the value of the decay constant in the
static limit \cite{SNH,ALLTON}:
\beq
f_B^{\infty} \simeq (1.98 \pm 0.31)f_\pi,
\eeq
the $1/M_b$-corrections are found to be \cite{SNA,SN4,NEU,ALLTON}:
\bea
&&f_B \sqrt{M_b} \simeq (0.33 \pm 0.06)\mbox{GeV}^{3/2}
\alpha_s^{1/{\beta_1}}
\Bigg\{ 1-\nnb\\&&\frac{2}{3}\frac{\alpha_s}{\pi}
-\frac{(A\simeq 1.1~ \mbox{GeV})}{M_b}+\frac{
(B\simeq  0.7~\mbox{GeV}^2)}{M^2_b}\Bigg\}
\eea
which,
one can $qualitatively$ compare with the one obtained from
the analytic expression of the moments or from the
semilocal duality sum rule
 leading to the $interpolating~formula$
\cite{ZAL}:
\bea
&&f_B \sqrt{M_b} \approx  \frac{E^{3/2}_c}{\pi\sqrt{2}}
\alpha_s^{1/{\beta_1}}
\ga \frac{M_b}{M_B} \dr^{3/2}\Bigg\{ 1-
\nnb\\
&&\frac{2}{3}\frac{\alpha_s}{\pi}
+\frac{3}{88}\frac{E^2_c}{M^2_b}
-\frac{\pi^2}{2} \frac{\la \bar uu \ra}{E^3_c}+...\Bigg\},
\eea
and gives for $E_c\simeq 1.3$ GeV:
\bea
A &\approx &\frac{3}{2}(M_B-M_b) \simeq 1~\mbox{GeV},\nnb \\
B &\approx& \frac{3}{88}E_c^2-\frac{9}{8}(M_B-M_b)^2 \simeq 0.5
{}~\mbox{GeV}^2,
\eea
\nin
 The $SU(3)$-breaking effects to these decay constants have
 also been estimated $analytically$ to be \cite{SN2}:
\beq
\frac{f_{D_s}}{f_D} \simeq
\frac{f_{B_s}}{f_B}\simeq (1.16 \pm 0.04)f_\pi~,
\eeq
which is in good agreement with the range of values obtained from
different
lattice groups \cite{LELLOUCH}.
The corresponding value of $f_{D_s} \simeq (1.55 \pm 0.10)f_\pi$
is still compatible (within the errors)
with the recent (indirect) measurements
from WA75 and CLEOII (see e.g.\cite{SCOTT}), which need to be
tested from direct
(though difficult) measurements of the leptonic widths.
However, independently of the
charm quark mass-value, which affects strongly the value of
$f_D$ \cite{SN2} ($f_D$ increases for decreasing $M_c$),
a value of $f_{D_s}$ larger than the {\it
rigorous} upper bound of 2.14$f_\pi$ deduced from \cite{SNZ}
and Eq. (18) is unlikely from the QSSR approach within the
standard SVZ-expansion.
\nin
We have also tested the validity
of the vacuum saturation  for the $B_B$-parameter, and we
found that the radiative corrections for the non-factorized
correlators are quite small (less than 15\%),
from which we deduce \cite{PIVO}:
\beq
 B_B \simeq 1 \pm 0.15.
\eeq
\section{Heavy-to-light transition-form factors}
\nin
One can extend the analysis done for the two-point correlator to
the
more complicated case of three-point function, in order to study
the form
factors related to the $B\rar \pi(\rho)l\nu$
and $B\rar K^*\gamma$ rare
decays. In so doing, one can consider the generic three-point
function:
\bea
V\equiv -\int d^4x~ d^4y ~e^{i(p'x-py)} ~\la 0|{\cal T}
 J_L(x)O(0)J^{\dagger}_B(y)|0\ra ,
\eea
where $J_L, ~J_B$ are the currents of the light and $B$ mesons;
$O$
 is the
weak operator specific for each process (penguin for the $K^*
\gamma$,
weak current for the semileptonic); $q \equiv p-p'$ is the
momentum transfer.
The vertex obeys a double dispersion relation, which can be
improved
into the $unique$ \footnote{The $popular$ double exponential sum
rule is not appropriate here as in this sum rule the OPE blows up
for
$M_b \rar\infty$.}
hybrid sum rule (HSR) \cite{SNA,SN5}:
\bea
{\cal H} (n, \tau') =\frac{1}{\pi^2}
\int_{M^2_b}^\infty \frac{ds}{s^{n+1}}
\int_0^\infty ds'~e^{-\tau' s'}~\mbox{Im}V(s,s'),
\eea
corresponding to a finite number $n~(n\approx 1-2$ in the present
processes) of
the moments for the heavy-quark
channel and to the Laplace for the light one.
We have
studied analytically the different form factors
\footnote{In the standard notations, the relevant form factors are
$f_+,~(A_1,~A_2,~V)$ for $B\rar\pi(\rho)l\nu$ and $F_1$ for $B\rar
K^*\gamma$ decays.} entering the previous
processes \cite{SN6},
and we found that they are dominated $universally$,
for $M_b \rar \infty$, by the
the light-quark-condensate contribution as:
\beq
F(0) \sim \frac{\la \bar dd \ra}{f_B}\aga 1+\frac{{\cal I}_F}
{M^2_b}\adr,
\eeq
where ${\cal I}_F$ is the integral from the perturbative triangle
graph, which is constant as $t'^2_cE_c/\la \bar dd \ra$ ($t'_c$ and
$E_c$ are the continuum thresholds of the light and $b$ quarks)
for large values of $M_b$. Unlike the case of $f_B$, where the
perturbative graph and the $\la \bar qq\ra$ condensate are of the
same order in $M_b$, the present dominance of
the $\la \bar qq\ra$ condensate allows a good separation of the
lowest ground state contribution from the radial excitation.
It also
indicates that at $q^2=0$ and to leading order in $1/M_b$,
all form factors behave like $\sqrt{M_b}$,
although, in most cases, the coefficients of $1/M_b$ due mainly
to $f_B$ and of $1/M^2_b$ due to the perturbative graph are large
\footnote{This feature also indicates that it is $dangerous$ to
extrapolate the $M_b$-dependence obtained at the $c$-quark mass to
higher quark mass values.}. In the particular
case of  $B\rar\pi l\nu$, the form
factor can be simply written to leading order \cite{DOSCH}:
\beq
f_+(0)\simeq \ga\frac{1}{4f_\pi}\dr\ga\frac{f^2_\pi}{f_B}\dr\simeq
0.15,
\eeq
to be compared with the numerical estimate 0.25 (the factor
$f^2_\pi$ reflects the off-shellness of the pion) and to the
pion coupling to hadron pairs \cite{SNB}.
The
study of the $q^2$ behaviours of the form factors shows
 that, with the
exception of the $A_1$ form factor, their $q^2$ dependence
is only due to the non-leading $1/M^2_b$ perturbative graph, so
that for $M_b \rar \infty$,
these  form factors remain constant from $q^2=0$ to $q^2_{max}$
and have a weaker $q^2$-dependence
( polynomial in
$q^2$ that can be resummed), than the pole model at finite $M_b$
(here the
value of pole mass which fits the form factors is about 5--6 GeV
\cite{BALL} (see also \cite{LEL})).
The resulting $M_b$ behaviour at $q^2_{max}$ is the one expected
from the heavy quark symmetry. The situation for the $A_1$ is
drastically different from
the other ones and from the pole parametrization.
Here the Wilson coefficient of the $\la \bar dd \ra$
condensate contains a $q^2$ dependence with a $wrong$ sign and
reads:
\beq
A_1(q^2) \sim \frac{\la \bar dd \ra}{f_B}\aga 1-\frac{q^2}{M^2_b}
\adr ,
\eeq
which, for $q^2_{max} \equiv (M_B-M_\rho)^2$, gives the expected
$1/\sqrt{M_b}$ behaviour. This result also explains the numerical
observation in \cite{BALL} (similar conclusions using alternative
approaches have also been reached in \cite{GOURD}).
Numerically, we obtain at $q^2=0$, the value of the $B\rar
\rho(K^*)\gamma$ form factors:
\bea
F_1^{B\rar\rho } \simeq 0.27 \pm 0.03,~~
\frac{F_1^{B\rar K^*}}{F_1^{ B\rar\rho}}\simeq 1.14 \pm 0.02,
\eea
which leads to the branching ratio $(4.5\pm 1.1)\times 10^{-5}$,
in perfect
agreement with the CLEO data, while the numerical agreement with
the estimate in \cite{ALI} from light cone sum rule (for a
criticism on the unreliability for the
construction of the hadronic wave functions on the light-cone,
see e.g. \cite{ECK})
may only be due to the importance of the perturbative
contribution at this scale \footnote{Lattice results on the
radiative and semi-leptonic decays are reported in \cite{BOUC}}.
For the semileptonic decays, a determination of the ratios of
the form factors gives a  more precise prediction \cite{SN5}
than from a direct estimate of the absolute values:
\bea
&&\frac{A_2(0)}{A_1(0)} \simeq \frac{V(0)}{A_1(0)}
\simeq 1.11 \pm 0.01,    \nnb  \\
&&\frac{A_1(0)}{F_1^{B\rar \rho}(0)} \simeq 1.18 \pm 0.06,~~
\frac{A_1(0)}{f_+(0)} \simeq 1.40 \pm 0.06.\eea
Combining these results with the ``world average" value of $f_+(0)=
0.25 \pm 0.02$ and the one of $F_1^{B\rar \rho}(0)$, one can
deduce the rate and polarization:
\bea
&&\Gamma_\pi \simeq (4.3\pm 0.7)
|V_{ub}|^2 \times 10^{12}~\mbox{s}^{-1}~~~
\frac{\Gamma_\rho }{\Gamma_\pi} \simeq 0.9 \pm 0.2 \nnb \\
&&\frac{\Gamma_+}{\Gamma_-} \simeq 0.20\pm 0.01~~~~~
\alpha \equiv 2\frac{\Gamma_L}{\Gamma_T}-1 \approx -0.6.
\eea
These precise results
may indicate that, $V_{ub}$ can be reached with a good accuracy
from the exclusive modes.
The non-pole behaviour of $A_1$ affects strongly the different
estimates in Eq. (27), in particular the ones of
$\Gamma_\rho /\Gamma_\pi$ and
$\alpha$, such that a firm prediction of these quantities needs a
an improved good control of the $q^2$-dependence of the
corresponding form factors.
We extend the previous analysis to the estimate of the $SU(3)$
breaking in the ratio of the form factors:
\beq
 R_P\equiv f_{+}^{P\rar K}(0) /f_{+}^{P\rar \pi}(0),
\eeq
where $P\equiv \bar B,~D$. Its analytic expression is given in
 \cite{SN8} and leads to the numerical result:
\beq
R_B = 1.007 \pm 0.020 ~~~~~~R_D=1.102\pm 0.007,
\eeq
which is typically of the same size
as the one of $f_{D_s}$ and of the $B\rar K^*\gamma$ discussed
before
and reinforces the credibility of the present estimate. However,
it is quite surprising that using the previous value of $R_D
$ into the present value of the CLEO data \cite{CLEO}:
\beq
\frac{Br(D^+\rar \pi^0 l\nu)}
{Br(D^+\rar \bar{K}^0 l\nu)}= (8.5 \pm 2.7 \pm 1.4)\%,
\eeq
one deduces:
\beq
V_{cd}/V_{cs}=0.322\pm 0.056,
\eeq
which is much larger than the value $0.226\pm 0.005$ derived from
the
unitarity of the CKM matrix. This $apparent$ discrepancy
needs a further measurement of the previous process before a firm
conclusion can be drawn (recall that MARKIII data \cite{MARK}
would imply a value $0.25 \pm 0.15$ compatible within the errors
with Eq. (31).).
\section{$B^*B\pi(\gamma)$ couplings and $D^*\rar D\pi(\gamma)$
widths}
As has been studied recently in \cite{DOSCH2}, the previous
processes are
very similar to the other heavy-to-heavy transitions as they are
dominated
by the perturbative graph contributions. The non-leading $1/M_b$
corrections
for the radiative decays are large as they come mainly from the
heavy quark
component of the electromagnetic current. This contribution is
essential for
explaining the large charge dependence in the observed radiative
decay widths.
For the $B^*$ meson, our predictions {\it without any free
parameters} are:
\beq
g_{B^*B\pi}\simeq 14 \pm 5~~~~~~~~\Gamma_{B^{*-}\rar B^-\gamma}/
\Gamma_{B^{*-}\rar B^-\gamma}\simeq 2.5,
\eeq
where the latter indicates a large isospin violation, which
deviates strongly
from the na\"\i ve static limit $e^2_u/e^2_d$ expectation,
therefore showing the importance of the $1/M_b$ corrections
in this channel. For
the $D^*$-one, we find:
\beq
\Gamma_{D^{*-}\rar D^0\pi^-}\simeq 1.54
\Gamma_{D^{*0}\rar D^0\pi^0}
\simeq (8\pm 5)~\mbox{keV},
\eeq
while:
\bea
\Gamma_{D^{*-}\rar D^-\gamma} &\simeq& (0.09^{+0.40}_{-0.07})~
\mbox{keV}\nnb\\
\Gamma_{D^{*0}\rar D^0\gamma} &\simeq& (3.7\pm 1.2)~\mbox{keV}.
\eea
The resulting total widths $\Gamma_{D^{*-}\rar all}\simeq (12\pm 7)
{}~\mbox{keV}$
and $\Gamma_{D^{*0}\rar all}\simeq (11\pm 4)~\mbox{keV}$ are much
smaller than
the present experimental upper limits. Improved measurements of
these widths
in the next $\tau$-charm factory machine should provide a decisive
 test of the predictions given here and should also help
to clarify the
disagreements among the
present theoretical predictions.
\section{Slope of the Isgur--Wise function and $V_{cb}$}
\nin
{}From the QSSR expression of the $universal$
Isgur--Wise function, to leading order in $1/M_b$ \cite{NEU}:
\bea
&&\zeta_{phys}(y\equiv vv')= \ga \frac{2}{1+y} \dr ^2
\Bigg \{
1 +\frac{\alpha_s}{\pi}
f(y) \nnb\\
&-&\la \bar dd \ra \tau^3 g(y)
+\la \alpha_sG^2 \ra \tau^4 h(y)
+g\la \bar dGd \ra \tau^5 k(y) \Bigg \},
\eea
where $\tau$ is the Laplace sum rule variable
and $f,~ h$ and $k$ are analytic functions of $y$. From this
expression, one
can derive the analytic form of the slope of the IW-function
\cite{SN7}:
\beq
\zeta'_{phys}(y=1) \simeq -1 + \delta_{pert} + \delta_{NP}
\simeq -1 \pm 0.02,
\eeq
where at the $\tau$-stability region:
\beq
\delta_{pert} \simeq -\delta_{NP} \simeq -0.04,
\eeq
which shows the near-cancellation of the non-leading corrections,
and we have added a generous $50 \%$
error of 0.02 for the correction terms. This result is in
agreement with
the improved bound of Taron--de Rafael on the slope of the
form factor
\cite{TARON}:
\beq
F'(vv'=1) \geq -1.5,
\eeq
based on the analyticity, the positivity
and a mapping technology of the elastic $b$-number form factor
$F$ defined as:
\beq
\la B(p')|\bar b \gamma^\mu b |B(b)\ra =(p+p')^\mu F(q^2),
\eeq
and normalized as $F(0)=1$
in the large mass limit $M_B \simeq M_D$.
The inclusion of the
 effects of the $\Upsilon$ states below $\bar BB$ thresholds
by using the sum of the $\Upsilon \bar BB$ couplings of
$0.34\pm 0.02$ from QSSR improves slightly
this bound to:
\beq
F'(vv'=1) \geq -1.34.
\eeq
Using the
relation of the form factor with the slope of the Isgur--Wise
function,
which differs by $-16/75 \log \alpha_s (M_b)$ \cite{FALK},
one can deduce the final bound:
\beq
\zeta'(1) \geq -1.04.
\eeq
The QSSR results are in good agreement with some
other significant estimates given in the literature.

\nin
 Let us now discuss the effects due to the
$1/M$ corrections, which can be done in two ways:
By calculating the $1/M_b$ corrections within HQET, \cite{NEUIW}
(resp. \cite{VAIN}) obtains:
\beq
{\eta}_A\zeta (1)=0.93\pm 0.03~(\mbox{resp}~0.89\pm 0.03)~.
\eeq
Here, the model-dependence enters when extrapolating the data at
$y=1$, and leads to:
\bea
V_{cb} \simeq
(39.9 \pm 2.9)\times 10^{-3}~(\mbox{resp}~(41.7 \pm 3)\times
10^{-3}),
\eea
Alternatively, one can use the value of the form factor at $q^2=0~
(y=1.5)$ from the sum rule in the {\it full theory} \cite{SN5}:
\beq
F(1.5)=0.53\pm 0.09
\eeq
and the data in the {\it whole range} of $y$ in order to
deduce the slope:
\beq
\rho^2\equiv -\zeta'\simeq 0.76\pm 0.2,
\eeq
from a linear parametrization of the form factor. The model
dependence
enters in this analysis through the curvature of the form factor.
The
main advantage of this approach is that it does not rely on the
previous
theoretical conflict at $y=1$. Using the different data from
CLEOII,
ARGUS and ALEPH \cite{ROSS}, we obtain the average:
\bea
V_{cb} \simeq
(39.9 \pm 1.2\pm 1.4)\times 10^{-3},
\eea
where the first error is from the data, and the second one is
from the
type of model-parametrizations. This result is in good agreement
with the
one in Eq. (43) and from the inclusive decays.
\section{Conclusion}
\nin
We have shortly presented different results from QCD spectral sum
rules
in the heavy-quark sector,
which are useful for further theoretical studies of the $B$-physics
and which complement the results from
alternative non-perturbative approaches.
{}From the experimental point of view,
QSSR predictions agree with available data,
but they also lead to some new features,
 which need to be tested in forthcoming
experiments.

\end{document}